\documentclass[12pt]{article}

\usepackage{amssymb}
\usepackage[dvips]{epsfig}
\usepackage[dvips]{graphics}
\usepackage{a4}

\begin{document}
\setlength{\unitlength}{1mm}
\title{On the Renormalizability of Theories with Gauge Anomalies}
\author{Rodolfo Casana$^1$, Sebasti\~ao A. Dias$^{1,2}$\\
 \it{\small $^1$Centro Brasileiro de Pesquisas F\'{\i}sicas}\\
\it{ \small Departamento de Campos e Part\'{\i}culas}\\ \it{\small
Rua Xavier Sigaud, 150, 22290-180, Rio de Janeiro, Brazil}\\
\it{\small $^2$Pontif\'\i cia Universidade Cat\'olica do Rio de
Janeiro}\\ {\it \small Departamento de F\'\i sica}\\ {\small \it
Rua Marqu\^es de S\~ao Vicente, 225, 22543-900, Rio de Janeiro,
Brazil}}
\date{}
\maketitle
\begin{abstract}
We consider the detailed renormalization of two (1+1)-dimensional
gauge theories which are quantized without preserving gauge
invariance: the chiral and the ``anomalous" Schwinger models. By
regularizing the non-perturbative divergences that appear in
fermionic Green's functions of both models, we show that the
``tree level" photon propagator is ill-defined, thus forcing one
to use the complete photon propagator in the loop expansion of
these functions. We perform the renormalization of these
divergences in both models to one loop level, defining it in a
consistent and semi-perturbative sense that we propose in this
paper.
\end{abstract}

\section{Introduction}
Cancellation of gauge anomalies is almost a {\it dogma} nowadays,
when one speaks about consistently defining a quantum gauge theory
\cite{Weinberg}. Traditional perturbation methods give no hope of
defining a finite and consistent theory in the presence of a gauge
anomaly, mainly because of the breakdown of Ward (or
Slavnov-Taylor) identities, which are extensively used to relate
counterterms and cancel divergent diagrams. It became a guiding
principle both for experimentalists (remember the search for the
top quark \cite{Cheng-Li}) and for theorists (for example, in
eliminating concurrent string theories \cite{Witten}).

In two dimensional space-time, however, indications appeared that
the situation could be more subtle. When we consider Weyl fermions
interacting minimally with an Abelian gauge field (the chiral
Schwinger model (CSM) \cite{JackiwR}), a consistent and unitary
quantum theory emerges in spite of the fact that gauge invariance
is lost). The resulting quantized model is dependent on a
parameter $a$ (Jackiw-Rajaraman parameter) which is introduced at
quantum level, and can not be fixed {\it a priori} to any value.
It is related to regularization ambiguities of the theory, that
can not be removed precisely because there is no ``natural"
regularization scheme (one that would preserve quantum gauge
symmetry).

A similar fact can occur within the context of the traditional
Schwinger model (Dirac fermions interacting minimally with an
Abelian gauge field \cite{Schwinger}). There exist regularizations
that preserve gauge invariance in all intermediate steps, and the
model, studied this way, provides very rich predictions about
several phenomena in Quantum Field Theory
\cite{Jackiwtop}-\cite{Manton}. However, if one chooses a
regularization that does not preserve explicitly gauge invariance
from the beginning \cite{Jackiw-Johnson}, one ends with an
effective action that is not gauge invariant, but depends on a
similar parameter $a$. When one sets $a=1$, one brings back the
desired gauge invariance. This is possible because there is no
true gauge anomaly in this case (gauge invariance can be restored
by the addiction of BRST co-homologically trivial counterterms).
In spite of that, even for values of $a$ different from one, the
theory is consistent and unitary, thus following closely what
happens in the CSM. This model, viewed in this context, is usually
called ``non-confining" Schwinger model \cite{Mitra-Rahaman}. As
the property of confinement depends crucially on the
renormalization of the model, thus not being firmly established up
to now, we prefer to call it ``anomalous" Schwinger model (ASM),
just to remember that gauge invariance is not explicitly preserved
by regularization.

In addiction to these facts, it has been discovered
\cite{Faddeev-Shatashvilli,Babelon-Schaposnik-Viallet,Harada-Tsutsui}
that there is a kind of symmetry restoration mechanism acting in
the background of anomalous gauge theories. This symmetry
restoration is possible thanks to the spontaneous appearance of a
new set of quantum degrees of freedom, the {\it Wess-Zumino
fields}, which, after being integrated along with the fermions,
help to build a gauge invariant effective action. This can be
explicitly demonstrated in (1+1) dimensions, for abelian gauge
theories, because the integration over the fermions furnishes a
quadratic integrated polynomial in $A_\mu$ \cite{Harada-Tsutsui}.
For higher dimensions or non-abelian theories this important
question remains opened, either because of the complexity of the
integration over the fermions \cite{Frolov-Slavnov} or of the
quite complicated interactions of the Wess-Zumino fields
\cite{Abdalla-Abdalla-Rothe}.

This {\it scenario} suggests that there is still a lot to be
learned about anomalous gauge theories, before discarding them
definitely (at least from a theoretical point of view). In fact,
even in the (1+1)-dimensional case there are unanswered questions.
Both in CSM and in ASM it is known that the fermionic Green's
functions are divergent (for any value of $a$, in the CSM
\cite{ChinesesPRD}, and for $a\neq 1$ in the ASM
\cite{Mitra-Rahaman}). It is readily seen that a fermionic wave
function renormalization is sufficient to turn both theories
finite \cite{Girotti,Boyanovsky}. Their renormalization, however,
is quite involved, and has never been performed in detail. The
main obstacle is that bosonization (necessary for solving the
models) is done in position space. It furnishes expressions quite
complicated to be Fourier transformed directly (remember that
renormalization is usually effected in momenta space). Also,
although the divergences are easily seen, they are not so easily
regularized, as their origin is not sufficiently clear.

It is our aim, in this paper, to solve these questions and
explicitly renormalize both models. We will show that it is
possible to do this considering an expansion of the fermion
propagator in terms of the full photon propagator (what we call a
``semi-perturbative" approach). This has to be preceded by a
careful analysis of the structure of the divergences and by
regularization, that we also do. We will end with well defined 1PI
functions to one loop (in the semi-perturbative expansion).

The paper is organized as follows: in section 2 we present a
calculation of the most relevant Green's functions and we show the
structure of the divergences. In section 3 we give a
regularization for them, in the gauge non-invariant formalism. In
section 4 we study the Ward identities and, in section 5, we
perform the renormalization of the model. Finally, in section 6,
we present our conclusions.
\section{Structure of Divergences}
The Schwinger model is defined by the following Lagrangian density
\footnote{Here $dx$
means $d^{2}x$. Our conventions are
\begin{eqnarray}
g_{\mu\nu}=\left(\begin{array}{cc} 1 &  0 \\
0 & -1 \\  \end{array}\right)\;=\;g^{\mu\nu}\quad,\quad
\epsilon^{\mu\nu}=\left(\begin{array}{cc} 0 &  1 \\ -1 & 0 \\
\end{array}\right)\;=\;-\epsilon_{\mu\nu}.\nonumber
\end{eqnarray}
\begin{eqnarray} \{\gamma^\mu,\gamma^\nu\}
\;=\;2\,g^{\mu\nu}&,&\gamma^\dag_0=\gamma_0\quad,
\quad\gamma^\dag_1=-\gamma_1\;.\nonumber
\end{eqnarray}
\begin{eqnarray}
\{\gamma_\mu,\gamma_5\}\;=\;0 &, & \gamma_5\;=\;\gamma_0
\gamma_1\quad,\quad\gamma^\dag_5\;=\gamma_5\;.\nonumber
\end{eqnarray}
\begin{eqnarray}
P_\pm=(1\pm\gamma_5)/2 &, & \tilde\partial_\mu=\epsilon_{\mu\nu}
\partial^\nu\;.\nonumber
\end{eqnarray}
}
\begin{eqnarray}  \label{lagV}
{\cal L}_{_V}[\psi,\overline\psi,A]=-\frac{1}{4}F_{\mu\nu}
 F^{\mu\nu} + \overline\psi(\;i\partial\!\!\!\slash
 +eA\!\!\!\slash\;)\psi .
\end{eqnarray}
The chiral Schwinger model is defined by
\begin{eqnarray} \label{lagC}
{\cal L}_{_C}[\psi,\overline\psi,A]=-\frac{1}{4}F_{\mu\nu}
F^{\mu\nu} + \overline\psi(\;i\partial\!\!\!\slash +eA\!\!\!\slash
P_+\;)\psi .
\end{eqnarray}
In both models, $\psi$ denotes a two dimensional Dirac fermion.
The effective action $W[A_\mu]$ is defined as
\begin{eqnarray}\label{aceft}
e^{iW[A_\mu]}=\int\!\!d\psi d\overline\psi\;\exp\left[\;i \int\!\!
dx\;\overline\psi\;iD[A_\mu]\;\psi \;\right] =\det iD[A_\mu]
\end{eqnarray}
where
\begin{equation}
iD[A_\mu]=\left\{\begin{array}{lcl}i\partial\!\!\!\slash
+eA\!\!\!\slash &,& \mbox{Schwinger model}\;,  \\
i\partial\!\!\!\slash +eA\!\!\!\slash P_+&, & \mbox{chiral
Schwinger model}\;.
\end{array}\right.
\end{equation}
We will calculate the fermionic determinant using a prescription
which is not gauge invariant~\cite{Jackiwtop,tese} for both
models. This prescription will be responsible for a value of $a$
different from $1$ in the vector case. We known that in the chiral
case, there is no gauge invariant prescription, whatever may be
the value of $a$.

The effective action in the vectorial case, $W_{_V}[A_\mu]$, is
given by
\begin{eqnarray}  \label{aceft1}
W_{_V}[A_\mu]=\int\!\!
dx\;\frac{1}{2}A_\mu(x)\left[m^2_{_V}\,g^{\mu\nu}
-\frac{e^2}{\pi} \frac{\partial^\mu\partial^\nu}{ \square }
\,\right]A_\nu(x),
\end{eqnarray}
where $m^2_{_V}$ is the mass dynamically
generated for the gauge field $A_\mu$,
\begin{eqnarray} \label{massaV}
m^2_{_V}=\frac{e^2}{2\pi}(a_{_V}+1).
\end{eqnarray}
In the chiral case we get the effective action $W_{_C}[A_\mu]$,
which is
\begin{equation} \label{acefq}
W_{_C}[A]=\frac{e^2}{8\pi}\int\!\!
dx\;A_\mu\left[\;a_{_C}g^{\mu\nu}- (\partial^\mu_x
+\tilde\partial^\mu_x)\frac{1}{\square} (\partial^\nu_x
+\tilde\partial^\nu_x )\right] A_\nu(x).
\end{equation}
In both cases the parameters $a_V$ and $a_C$  appear as a consequence
of ambiguities in the short-distance regularization of the fermionic
determinant.

The generating functional is
\begin{eqnarray}\label{gf0}
Z[\eta,\overline\eta,J]=N\!\!\int\!\!  dA_\mu d\psi d\overline\psi\;
\exp\left[\!i\! \int\!\!  dx\left({\cal L}[\psi,\overline\psi,A]
+\overline\eta\psi+\overline\psi\eta +J_\mu A^\mu\,\right)\right].
\end{eqnarray}
Now, we can integrate over the fermion fields and get
\begin{eqnarray} \label{gf1}
Z[\eta,\overline\eta,J]&\!\!=&\!\!\!\!\int\!\!
dA_\mu\;\exp\left[\,i \int\!\! dx\left(\frac{1}2
A_\mu\Gamma^{\mu\nu}A_\nu + J_\mu A^\mu \right)\right]\\ & &
\quad\quad\quad\times\exp\left[-i\int\!\! dx\,dy
\;\overline\eta(x)G(x,y;A)\eta(y)\,\right],\nonumber
\end{eqnarray}
where $\Gamma^{\mu\nu}$ is the 1PI two-point function of the gauge
field. For the vectorial coupling we have
\begin{equation}\label{DmunuV}
\Gamma^{\mu\nu}_{_V}= g^{\mu\nu} \left(\square+
m^2_{_V}\right) -\partial^\mu\partial^\nu- \frac{e^2}{\pi}
\frac{\partial^\mu\partial^\nu}{ \square } \;,
\end{equation}
while, in the chiral case,
\begin{equation}\label{DmunuC}
\Gamma^{\mu\nu}_{_C}= g^{\mu\nu}\left(\square
+\frac{a_{_C}e^2}{4\pi} \right) -\partial^\mu\partial^\nu -
\frac{e^2}{4\pi} \left(\partial^\mu+
\tilde\partial^\mu\right)\frac{1}{\square}
\left(\partial^\nu+\tilde\partial^\nu\right)
\end{equation}

The function $G(x,y;A)$ in (\ref{gf1}), is the two-point fermion
Green's function in the external field $A_\mu,\quad
D[A_\mu]G(x,y;A)= \delta(x-y)$. This Green's function can be exactly
computed in both models:
\begin{eqnarray}\label{grenvect}
G_{_V}(x,y;A) &=& \exp\left[-ie\int\!\! dz\;A_\mu(z)j^\mu_-(z,x,y)
\right]  P_-G_F(x-y) \;+ \quad\quad \\ & & \quad\quad+\;
\exp\left[-ie\int\!\! dz\;A_\mu(z)j^\mu_+(z,x,y)\;\right]
P_+G_F(x-y). \nonumber
\end{eqnarray}
\begin{equation} \label{greenqui}
G_{_C}(x,y;A)=\exp\left[-ie\int\!\!dz\;A_\mu(z)j^\mu_+(z,x,y)
\right] P_+G_F(x-y)+P_-G_F(x-y)
\end{equation}
Here, $G_F$ satisfies $i\partial\!\!\!\slash_x
G_F(x-y)=\delta(x-y)$ and $j^\mu_\pm$ is given by
\begin{eqnarray}
j^\mu_\pm (z,x,y)=(\partial^\mu_z \pm \tilde\partial^\mu_z)
[D_F(z-x) - D_F(z-y)]\;,
\end{eqnarray}
where $D_F(x)$, satisfies $\square D_F(x-y)=\delta(x-y)$.

Knowing this, we are able to compute the full photon and fermion
propagators. Besides, all the correlation functions of the theory
could, in principle, be exactly calculated in configuration space,
but not in momentum space, where one does not know how to bosonize
directly the theory.

The photon propagator $G_{\mu\nu}(x-y)=\langle 0|T A_\mu(x)
A_\nu(y) |0\rangle $ is straightforwardly computed from
(\ref{gf1}). When we consider the vectorial coupling, it yields
(in momentum space)
\begin{eqnarray}  \label{pfotonV}
i\,\tilde G^{^V}_{\mu\nu}(k)= \frac{1}{k^2-m^2_{_V}}
\left(g_{\mu\nu} -\frac{k_\mu k_\nu}{k^2}\right)-\frac{2
\pi}{e^2(a_{_V}-1)} \frac{ k_\mu k_\nu}{k^2}\;,
\end{eqnarray}
while, in the chiral case
\begin{equation}
i\tilde G^{^C}_{\mu\nu}(k)=\left[g_{\mu\nu}-\frac{k_\mu
k_\nu}{a_{_C}-1}\left(\frac{4\pi}{e^2}-\frac{2}{k^2}\right)+
\frac{k_\mu\tilde k_\nu +\tilde k_\mu
k_\nu}{(a_{_C}-1)\;k^2}\right]\frac{1}{k^2-m^2_{_C}},
\end{equation}
where $m^2_{_C}$ is defined as
\begin{equation} \label{masaC}
 m^2_{_C}=\frac{e^2 a^2_{_C}}{4\pi(a_{_C}-1)}.
\end{equation}
We observe that this propagator has a pole in $m^2_{_V}$
($m^2_{_C}$), that is, the photon acquires mass after the
quantization of the theory. We observe the explicit dependence of
the photon mass on $a_{_V}$ ($a_{_C}$), which leaves it
indefinite. Moreover, the photon propagator is divergence free.
Its high-momentum behavior is similar to the one in Proca's
theory.

Now we calculate the fermion propagator, $G(x-y)=\langle 0|
T\psi(x)\overline\psi(y)|0\rangle $. The vectorial fermion
propagator is
\begin{eqnarray} \label{exatfermV}
G_{_V}(x-y) &=& i\, \exp\left\{-\frac{4\pi\,i}{a^2_{_V}\!-1}\int\!\!
\frac{dk}{(2\pi)^2}\;\frac{1- e^{-ik{\cdotp}(x-y)}}{k^2}\right\}\\
& &\times\exp\left\{-\frac{2\pi\,i}{a_{_V}\!+1} \int\!\!
\frac{dk}{(2\pi)^2}\;\frac{1-e^{-ik{\cdotp}(x-y)}}{k^2-m^2_{_V}}
\right\}G_F(x-y)\,,
\nonumber
\end{eqnarray}
and the chiral fermion propagator is found to be
\begin{eqnarray} \label{exatfermC}
G_{_C}(x-y)&=&i\, \exp\left\{-\frac{4\pi\,i}{a_{_C}\!-1}\int\!\!
\frac{dk}{(2\pi)^2}\;\frac{1- e^{-ik{\cdotp}(x-y)}}{k^2-
m^2_{_C}}\right\}P_+G_F(x-y)\;+\nonumber \\ &  &\quad+\;i\,
P_-G_F(x-y)\,.
\end{eqnarray}

From (\ref{exatfermV}) and (\ref{exatfermC}), we see that the
fermionic propagators have an UV logarithmic divergence, but are
free of IR divergences.

This divergence is better understood  in momentum space. Although
we are not able to compute the Fourier transform exactly, we can
still write down the Schwinger-Dyson equation satisfied in the
vectorial case,
\begin{eqnarray}\label{sdysonV}
\left(\partial\!\!\!\slash_x
+e^2\int\!\!\frac{dk}{(2\pi)^2}\;f_{_V}(k) k\!\!\!\slash
\;e^{-ik\cdotp{(x-y)}}\right)G_{_V}(x-y)=\delta(x-y),
\end{eqnarray}
where $f_{_V}(k)$ is given by
\begin{eqnarray} \label{ffV}
f_{_V}(k)&=&-\frac{4\pi}{e^2(a^2_{_V}-1)}\;\frac{1}{k^2}-
\frac{2\pi}{e^2(a_{_V}+1)}\;\frac{1}{k^2-m^2_{_V}} \\
&=&-\frac{2\pi}{e^2(a_{_V}-1)}\;\frac{1}{k^2}-
\;\frac{1}{k^2(k^2-m^2_{_V})}.
\end{eqnarray}
The Schwinger-Dyson equations for the chiral case have to be written
separately for the right and left-handed fermions, as they have
different propagators. The right-handed part, $P_+G_{_C}=G^+_{_C}$,
satisfies
\begin{eqnarray}\label{sdysonC}
\left(\partial\!\!\!\slash_x +e^2\int\!\!\frac{dk}{(2\pi)^2}\;
f_{_C}(k) k\!\!\!\slash \;e^{-ik\cdotp{(x-y)}}\right)G^+_{_C}(x-y)
=P_-\delta(x-y),
\end{eqnarray}
and, for the left-handed part, $P_-G_{_C}=G^-_{_C}$, we obtain
\begin{equation}
\partial\!\!\!\slash_x G^-_{_C}(x-y)=P_+\delta(x-y).
\end{equation}
It denotes a free left-handed fermion. The function $f_{_C}$ is
given by
\begin{eqnarray} \label{ffC}
f_{_C}(k)&=&-\frac{4\pi}{e^2(a_{_C}-1)}\;\frac{1}{k^2-m^2_{_C}}.
\end{eqnarray}

We can express both equations, (\ref{sdysonV}) and (\ref{sdysonC}),
in a compact way
\begin{eqnarray}\label{sdyson}
\left(\partial\!\!\!\slash_x +e^2\int\!\!\frac{dk}{(2\pi)^2}\;
f(k) k\!\!\!\slash \;e^{-ik\cdotp{(x-y)}}\right)G(x-y)
=\delta(x-y)
\end{eqnarray}
This equation can be easily written in momentum space, where it
allows one to find a recursive equation for the fermion propagator
$\tilde G(p)$
\begin{eqnarray}\label{ecrcfer}
\tilde G(p)=\frac{i}{p\!\!\!\slash}-ie^2\int\!\!
\frac{dk}{(2\pi)^2}\;f(k)\,\frac{1}{p\!\!\!\slash}
\,k\!\!\!\slash\,\tilde G(p-k)\;.
\end{eqnarray}

The consequences of the previous equation will be studied in great
detail in one of the next sections. Now, we move to the
three-point Green's function, $G^\mu(x,y,z)=\langle 0
|T\psi(x)\overline\psi(y)A^\mu(z) |0\rangle$. In the vectorial
case, we get
\begin{eqnarray}\label{vertexV}
 G^\mu_{_V}(x,y,z)= i\,e\int\!\! \frac{dk}{(2\pi)^2}\;g^\mu_{_V}(k)
\left[\,e^{-ik\cdotp(z-x)}- e^{-ik\cdotp(z-y)}\,\right]
\,G_{_V}(x-y),
\end{eqnarray}
with $g^\mu_{_V}$ given by
\begin{eqnarray}
g^\mu_{_V}(k)=\frac{2\pi k^\mu}{e^2(a_{_V}-1)\,k^2} -
\frac{\gamma_5 \tilde k^\mu}{k^2\left(k^2-m^2_{_V}\right)}.
\end{eqnarray}
In momentum space $\tilde G^\mu_{_V}(p,-p-q,q)\equiv \tilde
G^\mu_{_V}(p,q)$
\begin{eqnarray}  \label{wigV}
\tilde G^\mu_{_V}(p,q)=i\,e \,g^\mu_{_V}(q)\left[\tilde
G_{_V}(p+q)-\tilde G_{_V}(p)\right].
\end{eqnarray}
The three-point function for the chiral case is
\begin{eqnarray}\label{vertexC}
 G^\mu_{_C}(x,y,z)= i\,e\int\!\! \frac{dk}{(2\pi)^2}\;g^\mu_{_C}(k)
\left[\,e^{-ik\cdotp(z-x)}- e^{-ik\cdotp(z-y)}\,\right]
\,G^+_{_C}(x-y),
\end{eqnarray}
with $g^\mu_{_C}$ given by
\begin{eqnarray}
g^\mu_{_C}(k)=\frac{4\pi}{e^2(a_{_C}-1)}\; \frac{k^\mu}{k^2-
m^2_{_C}} - \frac{a_{_C}}{a_{_C}-1}\;\frac{k^\mu+\tilde
k^\mu}{k^2\left(k^2-m^2_{_C}\right)}.
\end{eqnarray}
In momentum space $\tilde G^\mu_{_C}(p,-p-q,q)\equiv \tilde
G^\mu_{_C}(p,q)$
\begin{eqnarray}  \label{wigc}
\tilde G^\mu_{_C}(p,q)=i\,e \,g^\mu_{_C}(q)\left[\tilde
G^+_{_C}(p+q)-\tilde G^+_{_C}(p)\right].
\end{eqnarray}

We see that the divergence in this function is due to the
fermionic propagator. It can be easily seen that only Green's
functions with fermionic legs will have UV divergences
\cite{ChinesesPRD} in both models. A careful analysis leads us to
the conclusion that these UV divergences do not have perturbative
origin ~\cite{tiao1} as we are going to see in the next section.

We call attention to the fact that one can write the three-point
functions, in both theories, almost entirely in terms of the
two-point functions. This is certainly not an accident. Recently
Adam \cite{Adam2} and Radozycki and Namyslowski \cite{Radoz-Nami}
showed, in the context of the conventional (gauge invariantly
quantized) Schwinger model, that this property is linked to a
reduction of the infinite set of Schwinger-Dyson equations, so
that it is possible to express every $n$-point Green's functions
in terms of other ones, associated to lower $n$'s. In their paper,
Radozycki and Namyslowski used crucially the fact that gauge
invariance was preserved (in the form of a non-anomalous Ward
identity for the photon two-point function). The fact that the
same equations are still valid for anomalous models suggest that
this fact is not so much dependent on intermediate gauge
invariance as could seem at first sight. We will see that this
``factorization" is, in fact, a consequence of Ward identities
that are preserved also in the anomalous case, and is very
important for the renormalizability of the models.
\section{Regularization}
In this section we will begin by regularizing the vectorial
theory. We will use the point of view which is called {\it gauge
non-invariant formalism}~\cite{Abdalla-Abdalla-Rothe} where one
does not introduce a Wess-Zumino field to restore gauge symmetry
\cite{Faddeev-Shatashvilli,Harada-Tsutsui}. However, at least for
the case dealt with in this paper, the results are
coincident~\cite{tiao1}. In the gauge non-invariant formalism, the
vector field $A_\mu$ is decomposed in its longitudinal and
transverse parts, so
\begin{eqnarray}\label{decomp}
eA_\mu=\partial_\mu\rho-\tilde\partial_\mu\phi
\end{eqnarray}
If we substitute the decomposition (\ref{decomp}) in the action of
both models, we notice that, thanks to classical gauge invariance,
there is no kinetical term for the field $\rho$, that represents
the longitudinal part of $A_\mu$. We could try to decouple this
field, making a change of fermionic variables,
\begin{eqnarray} \label{measure}
\psi=e^{i\rho}\psi'&,& \overline\psi=\overline\psi' e^{-i\rho}.
\end{eqnarray}
However, the fermionic measure is not invariant under
(\ref{measure}), and changes  as
\begin{eqnarray}
d\psi d\overline\psi&=&d\psi'd\overline\psi' \;\exp\left[
\frac{i\,(a_{_V}-1)}{4\pi}\int\!\!
dx\;\partial_\mu\rho\partial^\mu\rho \,\right],
\end{eqnarray}
thus giving a kinetical term for the $\rho$ field. Putting into
the generating functional (\ref{gf0}), we obtain
\begin{eqnarray} \label{zregnic}
Z_{_V}[\eta,\overline\eta,J]&\!\!\!=&\!\!\!\int\!\!  d\rho d\phi d\psi
d\overline\psi\;\exp\left[i\int\!\!  dx\left(\;\frac{1}{2e^2} \phi
\square ^2 \phi +\overline\psi(i\partial\!\!\!\slash -
\tilde\partial\!\!\!\slash\phi )\psi+\right.\right.\\ &
&\left.\left.\!\!\!+ \frac{(a_{_V}-1)}{4\pi}\partial_\mu\rho
\partial^\mu\rho+\overline\eta e^{i\rho}\psi+\overline\psi
e^{-i\rho}\eta +\frac{1}{e}J_\mu
\partial^\mu\rho -\frac{1}{e}J_\mu
\tilde\partial^\mu\phi\;\right)\right].\nonumber
\end{eqnarray}
We notice that our tentative of decoupling was not successful, but
had the effect of generating a kinetical term to $\rho$. This
field is still coupled to the fermion fields, in a very
complicated way, through the fermionic sources. This is not a
usually noticed detail, as many people use to work with the
vacuum-vacuum amplitude, which has the sources set to zero. This
coupling induces the UV divergence. We will show this explicitly
for the vector case, the chiral case being completely analogous.
To compute the fermion two point function, one has to take two
functional derivatives of $Z$, given by equation (\ref{zregnic}),
with respect to $\eta$ and $\overline\eta$. Doing this and, in the
following, putting the sources to zero we obtain
\begin{eqnarray} \label{zregnic1}
0|T\psi(x)\overline\psi(y)|0\rangle &\!\!\!=&\!\!\!\int\!\!  d\rho
d\phi d\psi d\overline\psi\;\psi(x)\overline\psi(y)\;
\exp\left[i\int\!\!dx\left(\;\frac{1}{2e^2} \phi \square ^2 \phi
+\right.\right.\\ &
&\left.\left.\!\!\!+\overline\psi(i\partial\!\!\!\slash -
\tilde\partial\!\!\!\slash\phi )\psi+
\frac{(a_{_V}-1)}{4\pi}\partial_\mu\rho
\partial^\mu\rho+i\int\!\! dz\;
\rho(z) j(z,x,y)\right)\right],\nonumber \end{eqnarray} where
$j(z,x,y)=\delta(z-x)-\delta(z-y)$. We notice the quadratic
dependence of this expression in $\rho$, which enables its exact
integration. If one performs it naively, one finds a factor
\begin{equation}\exp\left\{-\frac{2\pi\,i}{a-1}
\int\!\!
\frac{dk}{(2\pi)^2}\;\frac{1-e^{-ik{\cdotp}(x-y)}}{k^2}\right\},
\end{equation}
which comes before the free fermion propagator, that is
logarithmically divergent. This divergence has an essentially
non-per\-tur\-ba\-ti\-ve nature, as can be seen by the
perturbative evaluation of the fermion two-point function (there
is no individual Feynman diagram inducing this divergence) and has
its parallel in the so called {\it gauge invariant formalism},
where one can see the same divergence appearing by the exact
integration over the Wess-Zumino fields \cite{tiao1}.

Having identified the origin of the divergence, the next step is
to regularize it. The {\it fundamental} observation here, is that
we could make everything finite if we had a better UV behavior for
the $\rho$ propagator. We can do this by means of Pauli-Villars
regularization. We add to the generating functional a new field
$\beta$ which has a large mass $\Lambda^2$ ~\cite{Zinn-Justin}
($\Lambda^2\rightarrow +\infty$). This field can be added
according to the usual recipe: a) add a term
\begin{equation}\label{PV1}\frac{(a_{_V}-1)}{4\pi}\beta(\square+
\Lambda^2)\beta\end{equation} to the action; b) modify the
interaction terms of $\rho$ through the substitution
$\rho\rightarrow\rho+\beta$; c) functionally integrate over
$\beta$ (immediately before this step, one performs the change of
variables in the $\rho$ integration $\rho^{\prime}=\rho-\beta$).
This defines the regularized generating functional
$Z^{^\Lambda}_{_V}[\eta,\overline \eta,J]$, which reduces to
$Z_{_V}[\eta,\overline\eta,J]$ in the limit of infinite $\Lambda$.
After the indicated manipulations into the regularized generating
functional, we get
\begin{eqnarray} \label{zregnic2}
Z^{^\Lambda}_{_V}[\eta,\overline\eta,J]&=&\!\!\!\int\!\!d\rho
d\phi d\psi d\overline\psi \,\exp\left[\,i\int\!\!
dx\left(\frac{1}{2e^2} \phi  \square ^2 \phi
+\overline\psi(i\partial\!\!\!\slash -
\tilde\partial\!\!\!\slash\phi )\psi\;+\right.\right.\\ &
&\hspace{-1.5cm}\left.\left.-\frac{(a_{_V}-1)}{4\pi\Lambda^2} \rho
\square ( \square + \Lambda^2)\rho+ \overline\eta
e^{i\rho}\psi+\overline\psi e^{-i\rho}\eta
+\frac{1}{e}J{\cdotp}\partial\rho-\frac{1}{e}J{\cdotp}
\tilde\partial\phi\; \right)\right].\nonumber
\end{eqnarray}
Now the propagator of the $\rho$ field has a better UV behavior
($k^{-4}$, for finite $\Lambda^2$). We are interested in the way
the original action changes after regularization. Then, we come
back to the original fields of the theory,
\begin{eqnarray}
\psi=e^{-i\rho}\psi'\quad,\quad \overline\psi=\overline\psi'
e^{i\rho}
\end{eqnarray}
(remembering that we have to take into account again the Jacobian)
and perform the inverse transformations $\rho=\frac{\partial_\mu
}{ \square } A^\mu\;,\; \phi=\frac{\tilde\partial_\mu}{ \square }
A^\mu$, to get the regularized generating functional in terms of
the original fields of the theory (\ref{lagV}),
\begin{eqnarray} \label{zreg22}
Z^{^\Lambda}_{_V}[\eta,\overline\eta,J] =\int\!\!  dA_\mu d\psi
d\overline\psi\;\exp\left(i\!\int\!\!  dx\;{\cal
L}^{^\Lambda}_{_V} [\psi,\overline\psi,A_\mu]+ J{\cdotp\!}A +
\overline\eta\psi +\overline\psi\eta\right)\;,
\end{eqnarray}
where ${\cal L}^{^\Lambda}_{_V}[\psi,\overline\psi,A_\mu]$ is the
Lagrangian density for the regularized theory
\begin{eqnarray}\label{lagregV}
{\cal L}^{^\Lambda}_{_V}[\psi,\overline\psi,A_\mu]
=-\frac{1}{4}F_{\mu\nu}F^{\mu\nu} +\overline\psi
(i\partial\!\!\!\slash+eA\!\!\!\slash) \psi-
\frac{e^2(a_{_V}-1)}{4\pi\Lambda^2} (\partial{\cdotp}A)^2.
\end{eqnarray}
We see a new term into the Lagrangian density, equivalent to a
``gauge fixing" condition with a gauge parameter which is $a$
dependent and proportional to $\Lambda^{-2}$. This new term allows
us to regularize the full theory.

Following the same steps we can regularize also the chiral
theory. We obtain
\begin{eqnarray}\label{lagregC}
{\cal L}^{^\Lambda}_{_C}[\psi,\overline\psi,A_\mu]
=-\frac{1}{4}F_{\mu\nu}F^{\mu\nu} +\overline\psi
(i\partial\!\!\!\slash+eA\!\!\!\slash\,P_+) \psi-
\frac{e^2(a_{_C}-1)}{8\pi\Lambda^2} (\partial{\cdotp}A)^2.
\end{eqnarray}
which also differs from the initial chiral Lagrangian density
(\ref{lagC}) by a ``gauge fixing" term, just like it happened in
the vectorial case.

In both cases, we should remember at this point that we do not
have the right to fix the gauge, as both theories are anomalous.
This is not what is being done. It is, in fact, quite curious that
the regularization can be performed in a way that is similar to a
gauge fixing.

From (\ref{zreg22}), we can now calculate the regularized Green's
functions. First, we find the Green's functions of the vectorial
case. The photon propagator is
\begin{eqnarray} \label{fdfdf}
i\,\tilde G^{^\Lambda}_{_V\mu\nu}(k)=\frac{g_{\mu\nu}}{k^2-
m^2_{_V}(a_{_V})}+f^{^\Lambda}_{_V}(k)\;k_\mu k_\nu
\end{eqnarray}
where the function $f^{^\Lambda}_{_V}$ is
\begin{equation}
f^{^\Lambda}_{_V}(k)=\frac{2\pi\Lambda^2}{e^2 (a_{_V}-1)}\;
 \frac{1}{k^2(k^2-\Lambda^2)}-\frac{1}{k^2(k^2-m^2_{_V})}
\end{equation}
which is finite when $\Lambda^2 \rightarrow +\infty$, as in the
non-regularized theory. The high momentum behavior of the
regularized photon propagator ($k^{-2}$) is better than in the non
regularized case. This allows the regularization of the fermion
propagator, as we will see.

The regularized fermion propagator is given by
\begin{eqnarray}
G^{^\Lambda}_{_V}(x-y)= i\,\exp\left\{ i\,e^2\int\!\!
\frac{dk}{(2\pi)^2}\;f^{^\Lambda}_{_V}(k) \left[ 1-
e^{-ik{\cdotp}(x-y)}\right]\right\}G_F(x-y),\nonumber
\end{eqnarray}
and satisfies the Schwinger-Dyson equation (\ref{sdyson}). In
momentum space the regularized fermion propagator $\tilde
G^{^\Lambda}_{_V}(p)$ also satisfies equation (\ref{ecrcfer}). A
power counting analysis of the terms above shows that the UV
logarithmic divergence is controlled. The regularized 1PI
two-point fermionic function is
\begin{eqnarray} \label{gam22fer1}
\tilde\Gamma^{^\Lambda}_{_V}(p)&=&p\!\!\!\slash  \left[\,1+
i\hbar\,e^2 \int\!\! \frac{dk}{(2\pi)^2} \;f^{^\Lambda}_{_V}(k)
\;k\!\!\!\slash\;\frac{1}{p\!\!\!\slash-k\!\!\!\slash} \;+\;{\cal
O}(\hbar^2)\;\right]
\end{eqnarray}
and the regularized three-point function
\begin{eqnarray} \label{verticexato2}
G^{^\Lambda\mu}_{_V}(x,y,z)=i\,e\int\!\!  \frac{dk}{(2\pi)^2}\;
g^{^\Lambda\mu}_{_V}(k)\left[\, e^{-ik\cdotp(z-x)}-
e^{-ik\cdotp(z-y)}\,\right]\,G^{^\Lambda}_{_V}(x-y)\;,
\end{eqnarray}
with $g^{^\Lambda\mu}_{_V}$ is given by
\begin{eqnarray}
g^{^\Lambda\mu}_{_V}(k)
=-\frac{2\pi\,\Lambda^2\,}{e^2(a_{_V}-1)}\; \frac{k^\mu}{k^2(k^2-
\Lambda^2)} - \frac{\gamma_5 \tilde k^\mu }{k^2(k^2 -m^2_{_V})}\;.
\end{eqnarray}
In momentum space $\tilde G^{^\Lambda\mu}_{_V}(p,-p-q,q)\equiv
\tilde G^{^\Lambda\mu}_{_V}(p,q)$ satisfies equation (\ref{wigV}).
And the regularized 1PI three-point function
$\tilde\Gamma^{^\Lambda\mu}_{_V} (p,q)$ is
\begin{eqnarray} \label{f3pt1}
\tilde\Gamma^{^\Lambda\mu}_{_V}(p,q)=e\,\frac{ \gamma^\mu
q\!\!\!\slash}{q^2} \left[\tilde \Gamma^{^\Lambda}_{_V}(p+q)-
\tilde \Gamma^{^\Lambda}_{_V}(p)\right].
\end{eqnarray}
So, we showed that the 1PI three-point function is regularized if
the 1PI two-point fermionic function is regularized, as expected.
It is easy to show, in a similar way, that all the fermionic Green's
functions are regularized too. Thus, we only need to renormalize the
fermion two-point function, as we will do in the next sections.

For the chiral case, the regularized Green's functions are also
easily computed. The regularized gauge field propagator is
\begin{eqnarray}
i\tilde G^{^\Lambda}_{_C\mu\nu}(k)&\!\!\!=&\!\!\!\frac{g_{\mu\nu}
\,(k^2- \Lambda^2) }{ (k^2-\omega_{_m})(k^2-\omega_{_\Lambda})}-
\frac{1}{a_{_C}-1}\;\frac{\Lambda^2\;(k_\mu \tilde k_\nu+\tilde
k_\mu k_\nu)}{k^2(k^2-\omega_{_m})(k^2- \omega_{_\Lambda})}+
\nonumber \\ & & \\ & & \hspace{-3cm} + \frac{ k_\mu
k_\nu}{k^2(k^2-\Lambda^2)}\left[ \frac{4\pi\Lambda^2}{e^2(a_{_C}
-1)} -\frac{(k^2- \Lambda^2)^2}{ (k^2-\omega_{_m})(k^2-
\omega_{_\Lambda})} - \frac{\Lambda^4/(a_{_C}-
1)^2}{(k^2-\omega_{_m})(k^2- \omega_{_\Lambda})}\right] ,\nonumber
\end{eqnarray}
where $\omega_{_m}$ and $\omega_{_\Lambda}$ satisfy
\begin{eqnarray}
\omega_{_\Lambda}+\omega_{_m}&=&\Lambda^2+\frac{e^2(a_{_C}+1)}{4\pi}
\nonumber\\ \omega_{_\Lambda}\omega_{_m}&=&\Lambda^2 m^2_{_C}\;.
\end{eqnarray}
We can solve these equations, to obtain
\begin{eqnarray}
\omega_{_\Lambda}&\approx& \Lambda^2-\frac{e^2}{4\pi(a_{_C}-1)}+
{\cal O}(\Lambda^{-2})\nonumber\\
\omega_{_m}&\approx&m^2_{_C}+{\cal O}(\Lambda^{-2})\;.
\end{eqnarray}

The regularized right-handed fermion propagator is given by
\begin{equation}\label{ferregC}
G^{^\Lambda+}_{_C}(x-y)=i\exp\left(ie^2\int\!\!\frac{dk}{(2\pi)^2}
\;f^{^\Lambda}_{_C}(k)\;\left[ 1-e^{-ik\cdotp(x-y)}\right]\right)
P_+\,G_F(x-y)
\end{equation}
and the left-handed fermion propagator is,
$G^{^\Lambda-}_{_C}(x-y)=iP_-\,G_F(x-y)$. The function
$f^{^\Lambda}_{_C}$ is defined as
\begin{eqnarray} \label{fegC}
f^{^\Lambda}_{_C}(k)&\!\!\!=&\!\!\!\frac{4\pi\Lambda^2}{e^2(a_{_C}
-1)}\, \frac{1}{k^2(k^2-\Lambda^2)} -\frac{\left[k^2- a_{_C}
\Lambda^2/(a_{_C}-1)\right]^2}{ k^2(k^2-\omega_{_m})(k^2-
\omega_{_\Lambda})(k^2-\Lambda^2)}.
\end{eqnarray}

The regularized 3-point function for the chiral case is
\begin{eqnarray}\label{vertexregC}
 G^{^\Lambda\mu}_{_C}(x,y,z)= i\,e\int\!\!
 \frac{dk}{(2\pi)^2}\;g^{^\Lambda\mu}_{_C}(k)
\left[\,e^{-ik\cdotp(z-x)}- e^{-ik\cdotp(z-y)}\,\right]
\,G^{^\Lambda+}_{_C}(x-y),
\end{eqnarray}
with $g^{^\Lambda\mu}_{_C}$ given by
\begin{eqnarray}
g^{^\Lambda\mu}_{_C}(k)=-k^\mu\,f^{^\Lambda}_{_C}(k)-
\frac{(k^\mu+\tilde k^\mu)\;\left[k^2- a_{_C}
\Lambda^2/(a_{_C}-1)\right]}{ k^2(k^2-\omega_{_m})(k^2-
\omega_{_\Lambda})}.
\end{eqnarray}
In momentum space $\tilde G^{^\Lambda\mu}_{_C}(p,-p-q,q)\equiv
\tilde G^{^\Lambda\mu}_{_C}(p,q)$ and we can write
\begin{eqnarray}  \label{wigcC}
\tilde G^{^\Lambda\mu}_{_C}(p,q)=i\,e
\,g^{^\Lambda\mu}_{_C}(q)\left[\tilde
G^{^\Lambda+}_{_C}(p+q)-\tilde G^{^\Lambda+}_{_C}(p)\right],
\end{eqnarray}
the regularized 1PI three-point function being
\begin{equation} \label{3pontosreg}
\tilde \Gamma^{^\Lambda\mu}_{_C}(p,q)=e\;\frac{\left(q^\mu+{\tilde
 q}^\mu\right)}{q^2}\left[P_-\,\tilde\Gamma^{^\Lambda}_{_C}(p+q)-
\tilde\Gamma^{^\Lambda}_{_C}(p)P_+\right]
\end{equation}

\section{Ward Identities}

In this section, we compute the Ward identities of both theories,
using the regularized Lagrangian densities
${\cal L}^{^\Lambda}_{_V}$ (eq. (\ref{lagregV})) and
${\cal L}^{^\Lambda}_{_C}$ (eq. (\ref{lagregC})).
As is usual, we perform the following gauge transformation
in the regularized generating functional, for the gauge field
\begin{equation}
A_\mu\rightarrow A_\mu +\frac{1}{e}\partial_\mu \lambda(x).
\end{equation}
The fermion transformation depends on the model. In the
vectorial case, it is given by
\begin{equation}\label{tfV}
\psi\rightarrow \psi +i\lambda(x)\psi\;,\quad \overline\psi
\rightarrow \overline\psi -i\lambda(x)\overline\psi ,
\end{equation}
whereas in the chiral case
\begin{equation}\label{tfC}
\psi\rightarrow \psi +i\lambda(x)P_+\psi\;,\quad \overline\psi
\rightarrow \overline\psi -i\lambda(x)\overline\psi P_- ,
\end{equation}
with infinitesimal $\lambda(x)$. In our framework, the fermionic
measure is not gauge invariant, changing as
$d\psi d\overline\psi\rightarrow J[A_\mu] d\psi d\overline\psi$.
$J$ is the Jacobian of the variable transformations (\ref{tfV}) or
(\ref{tfC}). The vectorial jacobian $J_{_V}$ is
\begin{equation}\label{JV}
\ln J_{_V}=-\frac{i\,(a_{_V}-1)}{2\pi}\int\!\!dx\;e\lambda\,
\partial\cdotp\!A +{\cal O}(\lambda^2),
\end{equation}
and the chiral jacobian is
\begin{equation}\label{JC}
\ln J_{_C}=-\frac{i}{4\pi}\int\!\!dx\;e\lambda\left[(a_{_C}-1)
\partial\cdotp\!A -\tilde\partial\cdotp A\right]\;+
{\cal O}(\lambda^2).
\end{equation}

Doing this, it is easy to obtain the fundamental Ward identity
satisfied by the generating functional of the 1PI functions
$\Gamma^{^\Lambda}[\psi,\overline\psi,A_\mu]$.
The Ward identity for the vectorial case, is
\begin{eqnarray} \label{WardIV}
i\;\frac{\delta \Gamma^{^\Lambda}_{_V}}{\delta\psi}\;\psi-
i\;\frac{\delta \Gamma^{^\Lambda}_{_V}}{\delta\overline\psi}
\;\overline\psi+ \frac{1}{e}\;\partial^\mu \;\frac{\delta
\Gamma^{^\Lambda}_{_V}}{\delta A^\mu}\;=\;\frac{e}{2\pi}
(a_{_V}-1)\left(1+ \frac{ \square }{\Lambda^2}\right)
\partial^\mu A_\mu,
\end{eqnarray}
while, for the chiral case,
\begin{eqnarray} \label{WardIC}
&&\hspace{-1cm}i\;\frac{\delta \Gamma^{^\Lambda}_{_C}}{\delta\psi}
\;P_+\,\psi- i\,P_-\,\frac{\delta \Gamma^{^\Lambda}_{_C}
}{\delta\overline\psi} \;\overline\psi+ \frac{1}{e} \;\partial^\mu
\;\frac{\delta \Gamma^{^\Lambda}_{_C}}{\delta A^\mu}
=\frac{e}{4\pi}\left[ (a_{_C}\!-1)\left(1+ \frac{ \square
}{\Lambda^2}\right) \partial^\mu
-\tilde\partial^\mu\right]A_\mu.\nonumber\\
\end{eqnarray}
So, the 1PI two-point bosonic function satisfies (in momentum space),
for the vectorial  case,
\begin{eqnarray}\label{WardfV}
k_\mu \tilde\Gamma^{^\Lambda\mu\nu}_{_V}(k)=\frac{e^2}{2\pi}
(a_{_V}-1) \left(1-\frac{k^2}{\Lambda^2}\right)k^\nu,
\end{eqnarray}
and for the chiral case
\begin{eqnarray}\label{WardfC}
k_\mu \tilde\Gamma^{^\Lambda\mu\nu}_{_C}(k)=\frac{e^2}{4\pi}\left[
(a_{_C}-1) \left(1-\frac{k^2}{\Lambda^2}\right)k^\nu-\tilde
k^\nu\right].
\end{eqnarray}
We see the non-transversality  of the photon propagator, which is
the sign of a gauge anomaly. Proceeding, we obtain another important
Ward identity,  involving the 1PI two-point fermionic function and
the 1PI three-point function. For the vector case we get
\begin{eqnarray} \label{Wardf1VV}
\frac{1}{e}\,q_\mu\,\tilde\Gamma^{^\Lambda\mu}_{_V} (p,q)=
\tilde\Gamma^{^\Lambda}_{_V}(p+q)-\tilde\Gamma^{^\Lambda}_{_V}(p)\;.
\end{eqnarray}
and for the chiral case
\begin{eqnarray} \label{WardfVC}
\frac{1}{e}\,q_\mu\,\tilde\Gamma^{^\Lambda\mu}_{_C} (p,q)=P_-
\tilde\Gamma^{^\Lambda}_{_C}(p+q)-\tilde\Gamma^{^\Lambda}_{_C}(p)P_+
\;.
\end{eqnarray}
 This identity can be obtained by direct
manipulation from equations (\ref{f3pt1}) and (\ref{3pontosreg})
respectively. It shows that we only need to renormalize the 1PI
two-point fermionic function. On the other hand, it repeats the
results already found by explicit computation of the three point
functions, thus establishing the origin of the previously
mentioned result. This Ward identity will be important for the
analysis of the renormalizability of both theories.


\section{Renormalization}

We will concentrate our analysis in the vectorial case, presenting
the results for the chiral case only at the end. In the usual way,
we express the regularized Lagrangian density (\ref{lagregV}) in
terms of renormalized quantities and their respective
renormalization constants
\begin{eqnarray} \label{lagbar}
{\cal L}^{^\Lambda}_{_V}&\!\!\!=&\!\!\!-\frac{1}{4}Z_A
\,F_{\mu\nu}F^{\mu\nu} - \frac{Z^2_e}{Z^2_\psi}
\frac{e^2(a_{_V}-1)}{4\pi\Lambda^2} (\partial_\mu
A^\mu)^2+Z_\psi\, \overline\psi i\partial\!\!\!\slash\psi  +
Z_e\,e A_\mu\overline\psi \gamma^\mu\psi\;. \nonumber \\
\end{eqnarray}
We define the bare fields $A^\mu_o$ and $\psi_o$, and the bare
coupling constant $e_o$ as
\begin{eqnarray}
A_o^\mu = \sqrt{Z_A}A^\mu \quad,\quad \psi_o = \sqrt{Z_\psi}\psi
\quad,\quad \label{cargbar} e_o = \frac{Z_e}{Z_\psi\,Z^{1/2}_A}
\,e\;.
\end{eqnarray}

The object of the renormalization procedure is to determine
$Z_\psi$, $Z_A$ and $Z_e$ that make  all Green's functions of
the theory finite. Possible ambiguities in the choice of these
constants, are parameterized through the imposition of
renormalization conditions.

The pure bosonic Green's functions do not have UV divergences.
Then, we do not need counter-terms to renormalize them, which
means
\begin{eqnarray} \label{zaza}
Z_A=1.
\end{eqnarray}

We remember the Ward identity (\ref{Wardf1VV}) satisfied by 1PI
bare functions
\begin{eqnarray}
\frac{1}{e_o}\,q_\mu\,\tilde\Gamma^{^\Lambda\mu}_{_V} (p,q)=
\tilde\Gamma^{^\Lambda}_{_V}(p+q)-\tilde\Gamma^{^\Lambda}_{_V}(p)\;.
\end{eqnarray}
Substituting in this equation the relation between the bare and
renormalized 1PI functions,
\begin{eqnarray}
\tilde\Gamma^{^R}_{_V}(p)&=&Z_\psi\,\tilde\Gamma^{^\Lambda}_{_V}(p)\\
\tilde\Gamma^{^R \mu}_{_V} (p,q)&=&Z_e\,
\tilde\Gamma^{^\Lambda\mu}_{_V}(p,q)  \; ,\label{gamre}
\end{eqnarray}
we obtain
\begin{eqnarray} \label{idwvt1}
\frac{1}{e}\,q_\mu\,\tilde\Gamma^{^R\mu}_{_V} (p,q)=
(Z_e\,Z^{-1}_\psi)^2
\left[\tilde\Gamma^{^R}_{_V}(p+q)-\tilde\Gamma^{^R}_{_V}(p)\right]\;.
\end{eqnarray}
On the other side, if we had started directly with the Lagrangian
written in terms of renormalized quantities, equation
(\ref{lagbar}),  we could {\it verify} that the renormalized
functions also satisfy the Ward identity
\begin{eqnarray} \label{idwvt2}
\frac{1}{e}\,q_\mu\,\tilde\Gamma^{^R\mu}_{_V}(p,q)=
Z_e\,Z^{-1}_\psi\left[
\tilde\Gamma^{^R}_{_V}(p+q)-\tilde\Gamma^{^R}_{_V}(p)\right]\;.
\end{eqnarray}
If we compare equations (\ref{idwvt1}) and (\ref{idwvt2}), we
obtain
\begin{eqnarray}
Z_e=Z_\psi\;.
\end{eqnarray}

Coming back to equation (\ref{cargbar}), and remembering that
$Z_A=1$, we see that
\begin{eqnarray} \label{ebare}
e_o=e\;.
\end{eqnarray}
We see that the coupling constant of the theory is not
renormalized, even when the theory is gauge non-invariantly
quantized. This shows that the universality of the electromagnetic
interaction, usually expressed by  $eA^\mu=e_o A^\mu_o$, can be
preserved into a gauge non-invariant renormalization scheme. In
particular, we see that the coupling constant will not depend on
the energy scale $\mu$ selected to impose the renormalization
conditions. Hence,  we have a null Callan-Symanzik beta function
\begin{eqnarray}
\beta=\mu\frac{\partial}{\partial\mu}e(\mu)\quad\Rightarrow\quad
\beta=0\;.
\end{eqnarray}
Analogous conclusions are reached also for the chiral case.

\subsection{Semi-perturbative analysis}

We start again from the Lagrangian density ${\cal
L}^{^\Lambda}_{_V} [\psi, \overline\psi,A_\mu] $ (\ref{lagregV}),
\begin{eqnarray} \label{laglamb1}
{\cal L}^{^\Lambda}_{_V}[\psi,\overline\psi,A_\mu] =-
\frac{1}{4}F_{\mu\nu} F^{\mu\nu}- \frac{e^2(a_{_V}-1)}{4\pi
\Lambda^2} (\partial{\cdotp}A)^2+\overline\psi
(i\partial\!\!\!\slash+eA\!\!\!\slash)\psi \;.
\end{eqnarray}
With a perturbative calculation in mind, we read the free photon
propagator
\begin{eqnarray} \label{probarf}
i\,\tilde G^{^0}_{\mu\nu}(k)= \frac{\,g_{\mu\nu}}{k^2}
+\frac{k_\mu k_\nu}{k^4}\left(\frac{2\pi
\Lambda^2}{e^2(a_{_V}-1)}-1\right) \;.
\end{eqnarray}
This propagator (\ref{probarf}) diverges quadratically
in the limit $\Lambda \rightarrow +\infty$. If we insert it
in the perturbative calculation of the correlation functions,
we will make the ultraviolet behavior of the individual graphics
worse in each consecutive perturbative order, implicating an
apparent non-renormalizability.

On the other hand, we can expand an equation analogous to
(\ref{ecrcfer}) for the regularized fermion propagator $\tilde
G^{^\Lambda}_{_V}(p)$ in powers of the function
$f^{^\Lambda}_{_V}$, and show that it is a series in $\hbar^n$, as
in the usual loop expansion. Writing $\hbar$ explicitly, we get
\begin{eqnarray}\label{expgfer}
\frac{1}{\hbar}\tilde
G^{^\Lambda}_{_V}(p)&=&\frac{i}{p\!\!\!\slash}\;+\;e^2\hbar
\int\!\! \frac{dk}{(2\pi)^2}\; f^{^\Lambda}_{_V}(k)\,
\frac{1}{p\!\!\!\slash}
\,k\!\!\!\slash\,\frac{1}{p\!\!\!\slash-k\!\!\!\slash}\;+\\ &
&\hspace{1cm}-\;ie^4\hbar^2 \int\!\!
\frac{dk}{(2\pi)^2}\frac{ds}{(2\pi)^2}
\;f^{^\Lambda}_{_V}(k)f^{^\Lambda}_{_V}(s)\frac{1}{p\!\!\!\slash}
k\!\!\!\slash \frac{1}{p\!\!\!\slash-k\!\!\!\slash}
s\!\!\!\slash\frac{1}{p\!\!\!\slash-k\!\!\!\slash-s\!\!\!\slash}\;
+\;\ldots\nonumber
\end{eqnarray}
In fact, we can show that expressions (\ref{expgfer}) and
(\ref{gam22fer1}) are equivalent to a loopwise expansion using the
exact photon propagator, and in this case a power
$(f_{_{V(C)}})^n$ corresponds to $\hbar^n$, or $n$ loops. Let us
take the second term in expression (\ref{expgfer}) and define
\begin{eqnarray}\label{g1} {\tilde{G}}_1(p;\Lambda)=\;e^2\hbar
\int\!\! \frac{dk}{(2\pi)^2}\; f^{^\Lambda}_{_V}(k)\,
\frac{1}{p\!\!\!\slash}
\,k\!\!\!\slash\,\frac{1}{p\!\!\!\slash-k\!\!\!\slash}\;.
\end{eqnarray}
We are going to show that \begin{eqnarray}\label{grf1}
\setlength{\unitlength}{1mm}
\hspace{-8.24cm}\scalebox{1}{\includegraphics[86,625][300,680]{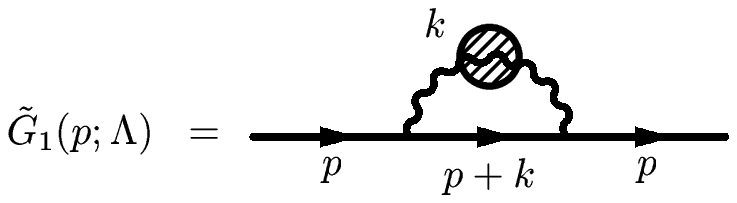}}
\end{eqnarray} where \begin{eqnarray}
\hspace{0cm}\scalebox{1}{\includegraphics[86,650][300,680]{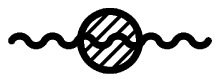}}
\nonumber
\end{eqnarray} denotes the regularized full photon propagator. The
graph in (\ref{grf1}) is expressed, in momenta space, as
\begin{eqnarray} \label{grf12} \tilde G_1(p;\Lambda)=e^2\int\!\!
\frac{dk}{(2\pi)^2}\;\frac{1}{p\!\!\!\slash}\gamma^\mu\frac{1}
{p\!\!\!\slash+k\!\!\!\slash}\gamma^\nu
\left(\frac{g_{\mu\nu}}{k^2-m^2(a)}+f^{^\Lambda}_{_V}(k)\,k_\mu
k_\nu\right)\frac{1}{p\!\!\!\slash} \end{eqnarray} The first term,
proportional to $g_{\mu\nu}$,  is cancelled because, in two
dimensions, $\gamma^\mu\gamma^\nu\gamma_\mu=0$. So,
\begin{eqnarray} \tilde
G_1(p;\Lambda)=e^2\int\!\!
\frac{dk}{(2\pi)^2}\;f^{^\Lambda}_{_V}(k)\,\frac{1}{p\!\!\!\slash}
k\!\!\!\slash\frac{1}{p\!\!\!\slash+k\!\!\!\slash}k\!\!\!\slash
\frac{1}{p\!\!\!\slash}\quad.\end{eqnarray} Using the
identity,\begin{eqnarray} \label{identity}
\frac{1}{p\!\!\!\slash}k\!\!\!\slash\frac{1}{p\!\!\!\slash+
k\!\!\!\slash}&=&\frac{1}{p\!\!\!\slash}-
\frac{1}{p\!\!\!\slash+k\!\!\!\slash} \end{eqnarray} and the fact
that $a\!\!\!\slash b\!\!\!\slash c\!\!\!\slash=c\!\!\!\slash
b\!\!\!\slash a\!\!\!\slash$, we obtain
\begin{eqnarray} \tilde G_1(p;\Lambda)=e^2\int\!\!
\frac{dk}{(2\pi)^2}\;f^{^\Lambda}_{_V}(k)\left(\frac{1}
{p\!\!\!\slash}k\!\!\!\slash\frac{1}{p\!\!\!\slash}-\frac{1}
{p\!\!\!\slash}k\!\!\!\slash\frac{1}{p\!\!\!\slash+k\!\!\!\slash}
\right).
\end{eqnarray} The first term inside parenthesis is cancelled
because it is the integral of an odd function of $k$. Changing $k$
by $-k$ we arrive where we wanted.

To proceed, let us define \begin{eqnarray} \tilde
G_{2}(p;\Lambda)= -ie^4\hbar^2\int\!\!
\frac{dk}{(2\pi)^2}\frac{ds}{(2\pi)^2}
\;f^{^\Lambda}_{_V}(k)f^{^\Lambda}_{_V}(s)\frac{1}{p\!\!\!\slash}
k\!\!\!\slash \frac{1}{p\!\!\!\slash-k\!\!\!\slash}
s\!\!\!\slash\frac{1}{p\!\!\!\slash-k\!\!\!\slash-s\!\!\!\slash}\;
\end{eqnarray}
We are going to show that, \begin{eqnarray} \label{2fermio}
\setlength{\unitlength}{1mm}
\hspace{-5cm}\scalebox{1}{\includegraphics[83,485][300,665]{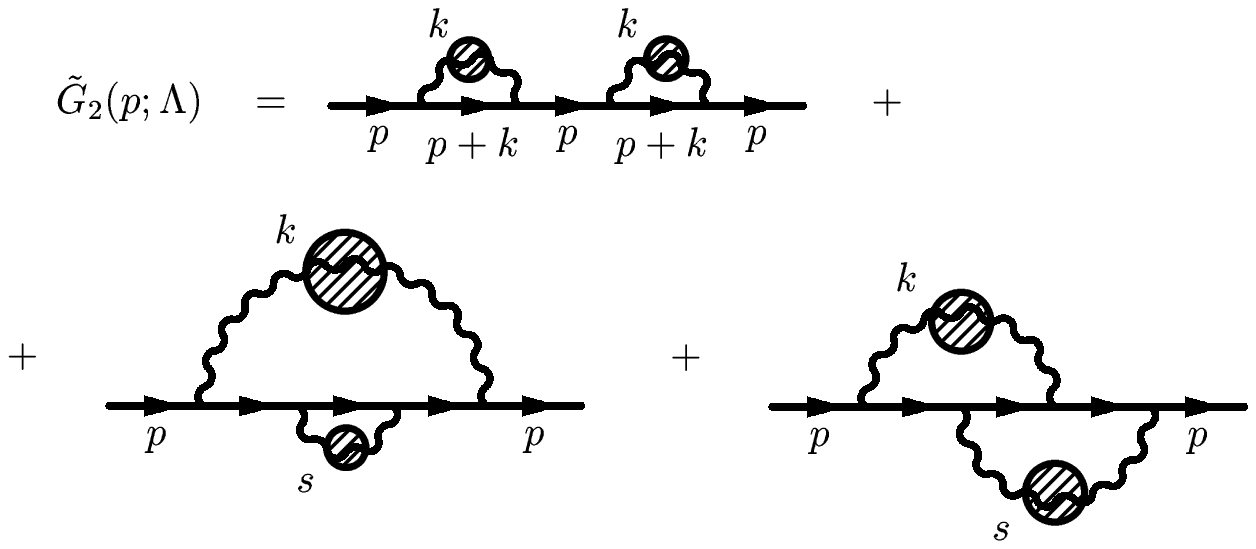}}
\end{eqnarray}
Computing the first graph in (\ref{2fermio}), we obtain
\begin{eqnarray} \label{21fermio} \setlength{\unitlength}{1mm}
\hspace{-3cm}\scalebox{1}{\includegraphics[86,625][300,680]{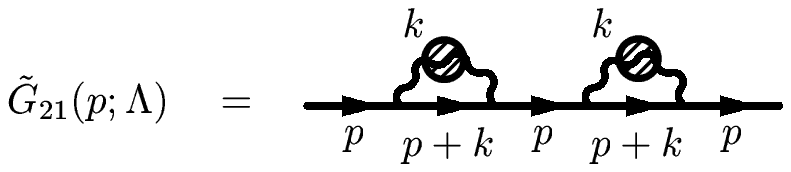}}
\end{eqnarray} \begin{eqnarray}
=-ie^4\int\!\!
\frac{dk}{(2\pi)^2}\frac{ds}{(2\pi)^2}\;f^{^\Lambda}_{_V}(k)
\,f^{^\Lambda}_{_V}(s)\;\frac{1}{p\!\!\!\slash}k\!\!\!\slash\frac{1}
{p\!\!\!\slash+k\!\!\!\slash}k\!\!\!\slash\frac{1}{p\!\!\!\slash}
s\!\!\!\slash\frac{1}{p\!\!\!\slash+s\!\!\!\slash}s\!\!\!\slash
\frac{1}{p\!\!\!\slash}
\end{eqnarray} \begin{eqnarray} \label{2fer1}
=-ie^4\int\!\!
\frac{dk}{(2\pi)^2}\frac{ds}{(2\pi)^2}\;f^{^\Lambda}_{_V}(k)
\,f^{^\Lambda}_{_V}(s)\,\left(
\frac{1}{p\!\!\!\slash}-\frac{1}{p\!\!\!\slash+k\!\!\!\slash}-
\frac{1}{p\!\!\!\slash+k\!\!\!\slash}k\!\!\!\slash\frac{1}
{p\!\!\!\slash+s\!\!\!\slash}\right),
\end{eqnarray} The second graph is, \begin{eqnarray} \label{22fermio}
\setlength{\unitlength}{1mm}
\hspace{-7cm}\scalebox{1}{\includegraphics[86,625][300,680]{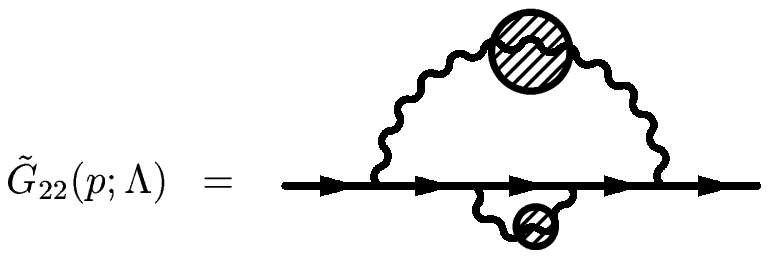}}
\end{eqnarray} \begin{eqnarray}
=-ie^4\int\!\!
\frac{dk}{(2\pi)^2}\frac{ds}{(2\pi)^2}\;f^{^\Lambda}_{_V}(k)
\,f^{^\Lambda}_{_V}(s)\;\frac{1}{p\!\!\!\slash}
k\!\!\!\slash\frac{1}{p\!\!\!\slash
+k\!\!\!\slash}s\!\!\!\slash\frac{1}{p\!\!\!\slash+k\!\!\!\slash+
s\!\!\!\slash}s\!\!\!\slash\frac{1}{p\!\!\!\slash+k\!\!\!\slash}
k\!\!\!\slash\frac{1}{p\!\!\!\slash}\nonumber \end{eqnarray}
\begin{eqnarray}\label{2fer2} =-ie^4\int\!\!
\frac{dk}{(2\pi)^2}\frac{ds}{(2\pi)^2}\;f^{^\Lambda}_{_V}(k)
\,f^{^\Lambda}_{_V}(s)
\,\left(
\frac{1}{p\!\!\!\slash}-\frac{1}{p\!\!\!\slash+k\!\!\!\slash}+
\frac{1}{p\!\!\!\slash} k\!\!\!\slash
\frac{1}{p\!\!\!\slash+k\!\!\!\slash+s\!\!\!\slash}
k\!\!\!\slash\frac{1}{p\!\!\!\slash}\right),\nonumber\\
& &
\end{eqnarray} The third one is, \begin{eqnarray} \label{23fermio}
\setlength{\unitlength}{1mm}
\hspace{-7cm}\scalebox{1}{\includegraphics[86,625][300,680]{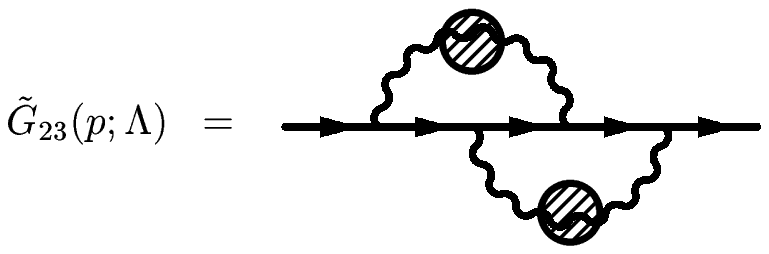}}
\end{eqnarray} \vspace{5mm}\begin{eqnarray} =-ie^4\int\!\! \frac{dk}
{(2\pi)^2}\frac{ds}{(2\pi)^2}\;f^{^\Lambda}_{_V}(k)
\,f^{^\Lambda}_{_V}(s)\;
\frac{1}{p\!\!\!\slash}k\!\!\!\slash\frac{1}{p\!\!\!\slash
+k\!\!\!\slash}s\!\!\!\slash\frac{1}{p\!\!\!\slash+k\!\!\!\slash+
s\!\!\!\slash}k\!\!\!\slash
\frac{1}{p\!\!\!\slash+s\!\!\!\slash}s\!\!\!\slash\frac{1}
{p\!\!\!\slash}\nonumber
\end{eqnarray}
\begin{eqnarray}\label{2fer3}
&\!\!=&\!\!-ie^4\int\!\!
\frac{dk}{(2\pi)^2}\frac{ds}{(2\pi)^2}\;f^{^\Lambda}_{_V}(k)
\,f^{^\Lambda}_{_V}(s)
\,\left(
-\frac{1}{p\!\!\!\slash}+\frac{1}{p\!\!\!\slash+k\!\!\!\slash}
\;+\right.\hspace{2.5cm}
\nonumber\\
&&\hspace{0.7cm}\left.-\;\frac{1}{p\!\!\!\slash}k\!\!\!\slash
\frac{1}{p\!\!\!\slash+k\!\!\!\slash+s\!\!\!\slash}
k\!\!\!\slash\frac{1}{p\!\!\!\slash}+
\frac{1}{p\!\!\!\slash+k\!\!\!\slash}k\!\!\!\slash\frac{1}
{p\!\!\!\slash+s\!\!\!\slash}-
\frac{1}{p\!\!\!\slash} k\!\!\!\slash
\frac{1}{p\!\!\!\slash+k\!\!\!\slash+s\!\!\!\slash}\right).
\end{eqnarray} Summing equations (\ref{2fer1}),(\ref{2fer2}) and
(\ref{2fer3}), we obtain
\begin{eqnarray} \tilde
G_2(p;\Lambda)=-ie^4\int\!\!
\frac{dk}{(2\pi)^2}\frac{ds}{(2\pi)^2}\;f^{^\Lambda}_{_V}(k)
\,f^{^\Lambda}_{_V}(s)\,
\left(\frac{1}{p\!\!\!\slash}-\frac{1}{p\!\!\!\slash+k\!\!\!\slash}
-\frac{1}{p\!\!\!\slash}
k\!\!\!\slash\frac{1}{p\!\!\!\slash+k\!\!\!\slash+s\!\!\!\slash}
\right).\end{eqnarray} Finally, identity (\ref{identity}) applied
twice gives us what we wanted to proof. The theorem is easily
established by finite induction and shown to be valid to all
orders. The generalization for the chiral Schwinger model is
straightforward.

This expansion of the exact fermion propagator in terms of the
exact photon propagator has been found previously in the
literature \cite{ChinesesPRD,Adam2,Radozycki}. However, the
fundamental role that it plays in the renormalization of
two-dimensional gauge theories has not been previously noticed or
used. A power counting analysis of expansion (\ref{expgfer}), with
$f^{^\Lambda}$ being $f^{^\Lambda}_{_V}$ or $f^{^\Lambda}_{_C}$,
shows the presence of a regularized UV logarithmic divergence (try
to compute the same graphs using $f$ instead of $f{^\Lambda}$) in
every loop that appears in the fermion propagator. This divergence
is similar to the one present in a Proca theory with fermions, due
to the bad high momentum behavior of the exact photon propagator.
However, it is possible to renormalize the theory in (1+1)
dimensions, because the bosonic sector is only quadratic in the
field $A_\mu$, after fermion integration. As we are going to see,
the renormalization of the fermionic sector can be done without
problems, once we recognize that we have to use the exact photon
propagator, instead of the tree level one.

The exact photon propagator itself does not exhibit the divergence
of the tree level one, which is cancelled when we add the terms of
the geometric sum that defines it. This shows that, in these
anomalous theories, the exact or complete photon propagator has to
be considered instead of the tree level one. The tree level of the
bosonic sector is thus indefinite. We call this kind of approach a
{\it semi-perturbative} one.

\subsection{Renormalization to 1-loop order}

Now, we can calculate the 1PI functions to 1-loop order (in this
semi-per\-tur\-ba\-ti\-ve sense) and impose renormalization
conditions, to determinate  the finite part of the renormalization
constants.

The regularized 1PI two point fermion function (\ref{gam22fer1})
$\tilde\Gamma^{^\Lambda}_{_V}$ to 1-loop order is
\begin{eqnarray}
\label{gam22} \tilde\Gamma^{^\Lambda}_{_V}(p)&\!\!\!=& \!\!\!
p\!\!\!\slash\left[\;1+\frac{\hbar}{2(a_{_V}-1)}\ln\left(1-
\frac{\Lambda^2}{p^2} \right)-\frac{\hbar}{2(a_{_V}+1)}\ln\left(1-
\frac{m^2_{_V}}{p^2} \right)\right]\end{eqnarray}

The renormalized $\tilde\Gamma^{^R}_{_V}$ function is given by
\begin{eqnarray}
\tilde\Gamma^{^R}_{_V}=Z^{^V}_\psi\,\tilde\Gamma^{^\Lambda}_{_V}\;,
\end{eqnarray}
where $Z^{^V}_\psi$ is the renormalization constant of the fermion
field. The next step, is the imposition of the renormalization
conditions that can fix the finite part of the renormalization
constants. This can be done by requiring that
\begin{eqnarray} \label{condtren}
\left.\tilde\Gamma^{^R}_{_V}(p)\right|_{p\!\!\!\slash=
\mu\!\!\!\slash}=\mu\!\!\!\slash\;.\label{condr1}
\end{eqnarray}
Using it, we get $Z^{^V}_\psi$ to 1-loop order,
\begin{eqnarray}\label{condr11}
Z^{^V}_\psi&\!\!\!=& \!\!\!
1-\frac{\hbar}{2(a_{_V}-1)}\ln\left(\,1-\frac{\Lambda^2}{\mu^2}
\,\right) +\frac{\hbar}{2(a_{_V}+1)}\ln\left(1-
\frac{m^2_{_V}}{\mu^2}\right) .
\end{eqnarray}
To this order, we get $\tilde\Gamma^{^R}_{_V}$ as
\begin{eqnarray} \label{gammareg}
\tilde\Gamma^{^R}_{_V}(p)&\!\!\!=&\!\!\!
p\!\!\!\slash\left[1+\frac{\hbar}{a^2_{_V}-1}\; \ln\left(
\frac{\mu^2}{p^2}\right)- \frac{\hbar}{2(a_{_V}+1)}\ln\left(
\frac{1-p^2/m^2_{_V}}{1- \mu^2/m^2_{_V}}\right)\right].\nonumber
\\ & &
\end{eqnarray}

The computation up to one loop of the 1PI 2-point function for the
chiral case will also be done. Defining the regularized
right-handed 1PI 2-point function as
$\tilde\Gamma^{^\Lambda+}_{_C}$, we obtain
\begin{eqnarray} \label{gam22fer1C}
\tilde\Gamma^{^\Lambda+}_{_C}(p)&=&p\!\!\!\slash P_+ \left[\,1+
i\hbar\,e^2 \int\!\! \frac{dk}{(2\pi)^2} \;f^{^\Lambda}_{_C}(k)
\;k\!\!\!\slash\;\frac{1}{p\!\!\!\slash-k\!\!\!\slash} \;+\;{\cal
O}(\hbar^2)\;\right] .
\end{eqnarray}
The function $ f^{^\Lambda}_{_C}$ can be written as
\begin{eqnarray}
f^{^\Lambda}_{_C}(k)&=&-\frac{\left[\omega_{_\Lambda}-a_{_C}
\Lambda^2/(a_{_C}-1)
\right]^2}{\omega_{_\Lambda}(\omega_{_\Lambda}-\omega_{_m})}
\frac{1}{(k^2-\Lambda^2)(k^2-\omega_{_\Lambda})}+\\ &
&\quad+\frac{\left[\omega_{_m}-a_{_C}\Lambda^2/(a_{_C}-1)
\right]^2}{\omega_{_m}(\omega_{_\Lambda}-\omega_{_m})}
\frac{1}{(k^2-\Lambda^2)(k^2-\omega_{_m})}.\nonumber
\end{eqnarray}

We compute the integral in (\ref{gam22fer1C}) to obtain
\begin{eqnarray} \label{gam22ferC}
\tilde\Gamma^{^\Lambda+}_{_C}(p)&=&p\!\!\!\slash P_+ \left[\,1-
\frac{\hbar}{a-1}\ln\left(\frac{\Lambda^2}{m^2_{_C}}\right)+
\frac{\hbar}{a-1}\ln\left(1-\frac{p^2}{m^2_{_C}}\right)\right]
\end{eqnarray}
Now, we renormalize this 1PI function $\tilde\Gamma^{^R+}_{_C}
=Z^{^C}_\psi\tilde\Gamma^{^\Lambda+}_{_C}$ and  use the
renormalization condition (\ref{condtren}) to get the one loop
wave function renormalization constant
\begin{eqnarray}\label{ZpsiC}
Z^{^C}_\psi=
1+\frac{\hbar}{a-1}\ln\left(\frac{\Lambda^2}{m^2_{_C}}
\right)-\frac{\hbar}{a-1}\ln\left(1-\frac{\mu^2}{m^2_{_C}}\right)
.
\end{eqnarray}
Then the renormalized 1PI 2-point function to one loop order is
\begin{eqnarray} \label{gamFerC}
\tilde\Gamma^{^R+}_{_C}(p)&=&p\!\!\!\slash P_+ \left[\,1+
\frac{\hbar}{a_{_C}-1}\ln\left(\frac{1-p^2/m^2_{_C}}{1-\mu^2/m^2_{_C}}
\right)\right] .
\end{eqnarray}


\section{Conclusions}

We have seen, through the examples of the vector and chiral
Schwinger models, how to renormalize an anomalous gauge theory, at
least in two dimensions. The main feature is that the theory is
renormalizable, in the usual sense, if the complete photon
propagator can be computed. This could be a good starting point to
attack the same question in four dimensions, if we could estimate
or take into account the main characteristics of the exact photon
propagator.

In the regularized version of the theory, the dependence on $a$ is
completely contained in the cut-off dependent ``gauge fixing"
parameter. If we were dealing with a ``normal" gauge theory, this
would suggest that the complete theory (that one where the gauge
field is also quantized) is independent of $a$, thus allowing one
to choose whether one keeps or not explicit gauge invariance in
any intermediate steps of the quantization. However, we must be
careful in analysing this, because of the apparently non-gauge
invariant nature of the quantum theory. It is necessary to
remember here that we are not fixing the gauge (we do not have the
right to do that, it is an anomalous theory), but regularizing
divergences of non-perturbative nature.

Finally, we call attention to the fact that, from the point of
view of a semi-perturbative renormalization, one does not
distinguish between a truly anomalous gauge theory and a
non-anomalous. It is only when the gauge field is fully quantized
that one can see this clearly, what is not usually done. Most
works limit themselves to integrating over the fermion fields and,
forgetting fermionic sources, manipulating a quadratic bosonized
theory. For many applications, this is not wrong. But, as we have
shown, many interesting aspects of the theory are revealed only
with the complete quantization.

We are currently investigating the physical consequences of these
renormalized versions of the theories. We can say, preliminarily,
that the parameter $a$ is apparently controlling the screening and
confinement properties. Progress in this direction will be
reported elsewhere.

\section{Acknowledgment}

This work was supported by Conselho Nacional de Desenvolvimento
Cient\'{\i}fico e Tecnol\'ogico, CNPq, Brazil.


\end{document}